\documentclass[prd,letterpaper,tightenlines,nofootinbib,twocolumn,floatfix,showpacs]{revtex4}

\usepackage{amssymb,graphicx,amsmath,color,float,soul}

\begin{document} 
\newcommand{\PDFFIG}{}
\title{Diffusion and escape times in the open-leaky standard map}
\author{L.~Lugosi$^{1}$, T.~Kov\'acs$^{1}$}
\affiliation{$^1$ E\"otv\"os Lor\'and University, Institute of Physics, H-1117 Budapest, P\'azm\'any P. s. 1/A, Hungary}

\begin{abstract}
We study the connection between transport phenomenon and escape rate statistics in two-dimensional standard map. For the purpose of having an open phase space, we let the momentum co-ordinate vary freely and restrict only angle with periodic boundary condition. We also define a pair of artificial holes placed symmetrically along the momentum axis where the particles might leave the system. As a consequence of the leaks the diffusion can be analysed making use of only the ensemble of survived particles. We present how the diffusion coefficient depends on the size and position of the  escape regions. Since the accelerator modes and, thus, the diffusion are strongly related to the system's control parameter, we also investigate effects of the perturbation strength. Numerical simulations show that the short-time escape statistics does not follow the well-known exponential decay especially for large values of perturbation parameters. The analysis of the escape direction also supports this picture as a significant amount of particles skip the leaks and leave the system just after a longtime excursion in the remote zones of the phase space.  
\end{abstract}

\date{\today}
\maketitle

\section{Introduction}
Fermi acceleration (FA) has been known since Fermi introduced \cite{Fermi1949} the concept to describe the high energy cosmic rays. In brief, particles gaining kinetic energy in unlimited amount caused by oscillating magnetic fields undergo different kind of diffusion in phase space \cite{Lichtenberg1983,Leonel2009,Shah2010,Gelfreich2011,Livorati2012,Kroetz2015}.

Since then numerous studies dealt with accelerator modes (AM) identified as regular islands in simple conservative dynamical systems as the motive of anomalous transport \cite{Korabel2007,Meiss2015,Zumofen1994,Zumofen1999,Benkadda1994,Harsoula,Lichtenberg1983,Ichikawa1989,Manos2013}. The paradigmatic example of these investigations is the two-dimensional (2D) standard map (SM) \cite{Chirikov} wherein one or both phase space coordinate is not bounded by any restriction. That is, the system is open and irregular motion might appear in the motion along that direction(s). In general case, normal diffusion describes the spreading of (initially nearby) particles in the phase space. 
Accelerator modes (usually quite tiny subsets of phase space) can keep chaotic orbits around themselves for shorter or longer times. As an impact these orbits seem to be regular during the trapping time. This effect of the AMs is called stickiness, one of the backbones of conservative dynamics \cite{Zaslavsky2007,Contopoulos2004,Lai2011,Meiss2015}. If stickiness is present determining the transport phenomenon might be problematic. Recently, global and local diffusion have been studied in standard map with and without AMs \cite{Harsoula}. The authors report that the global diffusion is normal when no accelerator modes present in the phase space and anomalous when accelerator modes exist. However, the local diffusion strongly depends on the initial conditions. They also claim that the diffusion turns to be normal after finite time even for initial conditions in the extreme sticky region.

One of the possible physical consequences of FA is that the particle takes so much kinetic energy that it can leave the system. One experimental realization of this action is to set a maximum value of a coordinate above which particles are outside the system. Livorati et al.~\cite{Livorati} applied this framework for the bouncing ball model in order to scrutinize the normal-ballistic diffusion transition in a one-dimensional.
Other implementation of particle escape is by defining artificial hole in the phase space of an originally closed system. ~\cite{Altmann,TT_leaky,Earth_Moon}. In this scheme as soon as a particle reaches the area of the leak, it is forced to leave the system and additional iterations will not be taken with that. Both aspects have profound theoretical background in the literature and wide-spread application among many disciplines.

The motivation of present work is to understand the role of vanishing particles in calculation of diffusion coefficient. To integrate these two concepts we use the unbounded or open SM in combination with finite size leaks in the phase space. We refer to this setup as \textit{open-leaky SM.}  Since the diffusion is strongly related to the non-linearity in SM that also governs the particles' dynamics, we focus on escape statistics in open-leaky system which is also connected to strength of periodic perturbation of the system.

The paper is organized as follows. In Sec.~\ref{model}, we present the model and details of the simulations and parameter choices. Sec.~\ref{result} is devoted to results about the transport and escape dynamics. In last section,  Sec.~\ref{conclusion}, we summarize what we have learned about the behavior of the open-leaky standard map. 

\section{Model and simulations}
\label{model}
The model we investigate is the well-known 2D chaotic standard map~\cite{Chirikov}. It is simple in a way that the phase space is only two dimensional but complex because of reach dynamics that might appear.
The defining equations are the following
\begin{equation}
\label{STM}
\begin{split}
p_{n+1} &=p_n\, +\frac{K}{2\,\pi} \,\sin(2\,\pi\,\Theta_n),\\
\Theta_{n+1} &=\Theta_n\, + p_{n+1} \qquad (\mathrm{mod}\, 1),	
\end{split}
\end{equation}
where $p $ and $\Theta$ are the momentum and the angle coordinates, $K$ is the non-linearity parameter, $n$ refers to the number of iterations. In present study, we only have \textit{mod 1} restriction for the angular coordinate which implies that the map is open along the $p$ axis letting the momentum reaches arbitrary large values. 

\begin{figure}[htb]
    \centering
    \includegraphics[width=\columnwidth] {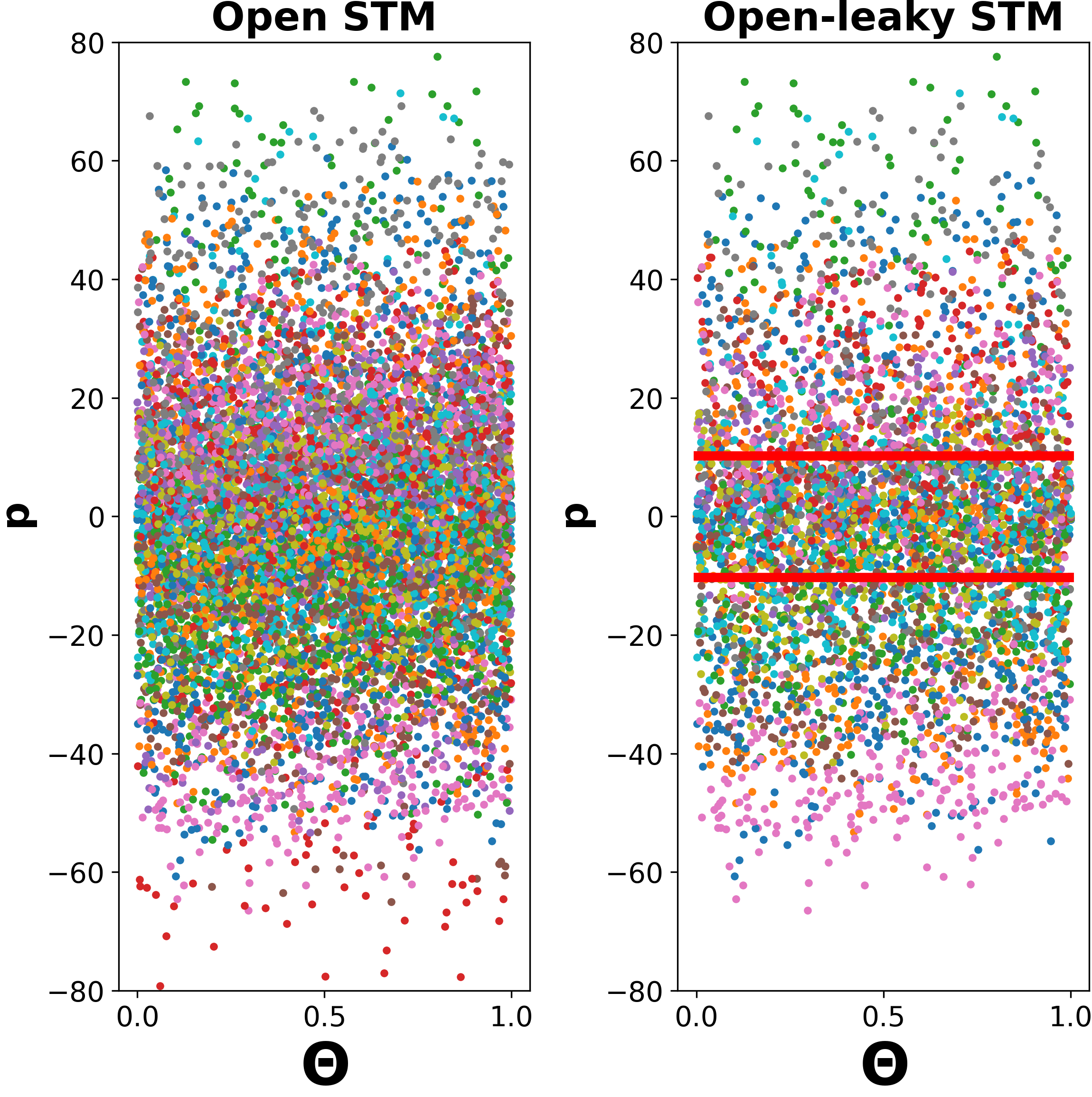}
    \caption{Trajectories of $100$ particles for $100$ iteration steps in open (left) and open-leaky (right) standard map with $K=28$. Note the asymmetry of the distribution in the right panel as a consequences of disappearing particles.}
	\label{phase_space}
\end{figure}

Fig.~\ref{phase_space} demonstrates a representative phase portrait of the open standard map with $K=28$ and the open-leaky layout with a pair of leaks (horizontal red lines) at $p=[10,10.5]$ and $[-10.5,-10]$. Here, the simulation includes only $100$ particles up to $100$ iteration steps in order to avoid overcrowding the plot. Colors refer to positions of the same particle at different time instant.

Accelerator modes, as special kind of initial conditions belonging to regular islands, can appear in the system \cite{Meiss1992}. These orbits look like ''leap frogs'' in  phase space as they jump by integer values along the $p$ axis as iteration goes on. Nevertheless, the appearance of period-1 AMs as a function
of the non-linearity parameter can be defined with the inequality~\cite{Chirikov}
\begin{equation}
2\pi l \leq K \leq \sqrt{(2\pi l)^2+16},
\label{acc_eq}
\end{equation}
here $l$ denotes a positive integer number, $K$ is as the same as in Eq.~\eqref{STM}. One can also obtain that the width of the successive $\Delta K$ intervals decreases by $\sim9/K$ \cite{Harsoula}. 

The diffusion coefficient is the expectation value of the squared distance for an ensemble of trajectories. It is determined numerically by the equation\footnote{In case of general 2D normal diffusion both coordinates have to be take into account. However, the difference in our simulations was less than $10^{-3}\,\%$. Consequently we used Eq.~\eqref{diffeq} to determine the coefficient.}~\cite{Harsoula}
\begin{equation}
\label{diffeq}
D(K)=\frac{\langle(p-p_0)^2\rangle}{n},
\end{equation}
where $D$ is the diffusion coefficient for a given $K$ value, $p_0$ denotes the initial momenta, and $\langle\dots \rangle$ refers to the average over the ensemble of trajectories. For completeness we note that $n$ should be large enough in order to get a saturation in $D.$ The $K$ dependence on the right hand side of Eq.~\ref{diffeq} is included in the calculation of $p$ coordinate according to Eq.~\ref{STM}. To demonstrate the facts above, we reproduce the classical spike structure of the $D(K)$ curve \cite{Harsoula,Manos2013} in Figure~\ref{deltaD} (yellow line).

It is also known that accelerator mode islands help the particles to take larger steps in one iteration, as a result, those particles can travel further in the phase space showing anomalous diffusion rather than normal for long timescale \cite{Ichikawa1989,Zaslavsky1997b,Rom-Kedar1999,Contopoulos2005,Manos2013}. Furthermore, orbits inside the AMs are ballistically transported in both directions $p\to\pm\infty.$

In physical systems when the total mechanical energy becomes positive, the equipotential surfaces open up, particles' orbit becomes unbounded and, therefore, can escape a pre-defined region. Similar phenomenon in open standard map wherein the cylinder-like phase space is extended to infinity can be observed at critical value of $K_C\approx 0.976.$ Beyond this value the last KAM-torus destroys and trajectories originating from chaotic sea can visit the entire phase space.

In order to be able to observe escape in open standard map we have two choices. First, by defining a maximum value of momentum $p_{\mathrm{max}.}$ In this case particles leave the system, and do not contribute to the dynamics at all, if the condition $p>p_{\mathrm{max}}$ is fulfilled. This setup can be thought of as an open flow in the phase space with a bounded region ($\pm p_{\mathrm{max}}$ in our case) where irregular motion appears before they quit this domain and complexity ceases. We recall that this phenomenon has been extensively studied and can be described by chaotic scattering \cite{Ott2002,Tel2005,Lai2011}.

Second, closed systems can be open-up artificially by placing a hole in the phase space and let the particles leave the system through this hole, see the thorough review by \cite{Altmann}. It can also be shown that the Poincar\'e recurrences in closed chaotic systems corresponds to escape through a leak provided the leak is positioned exactly to the recurrence region \cite{Altmann2009}.

Both cases above show paradigmatic escape statistics of survived particles either through the pre-defined border \cite{Contopoulos2004,Schelin2009,Gouillart2011} of the system or through a leak \cite{Bleher1988,Schneider2002,Sanjuan2003,Nagler204,Kovacs}. That is, escape of particles in fully hyperbolic dynamics and for short times in weakly chaotic systems shows exponential decay. Moreover, the number of surviving trajectories in weakly chaotic systems for long timescales follows a power-law distribution.

In order to examine transport and escape in the extended standard map we place two holes at predefined positions in the phase space. In terms of its size, we utilize a pair of leaks along the whole $\theta=[0,1]$ interval with varying height along $p$. We are interested in the diffusion along the $p$ direction.

In most of our simulations scrutinizing the diffusion coefficient we put $400$ particles initially on a lattice at $p=[-0.5,0.5]$ and $\Theta=[0,1]$ intervals. The length of iteration is $500$.

\section{Results}
\label{result}
\subsection{Diffusion}

\begin{figure}[htb]
    \centering
    \includegraphics[width=\columnwidth] {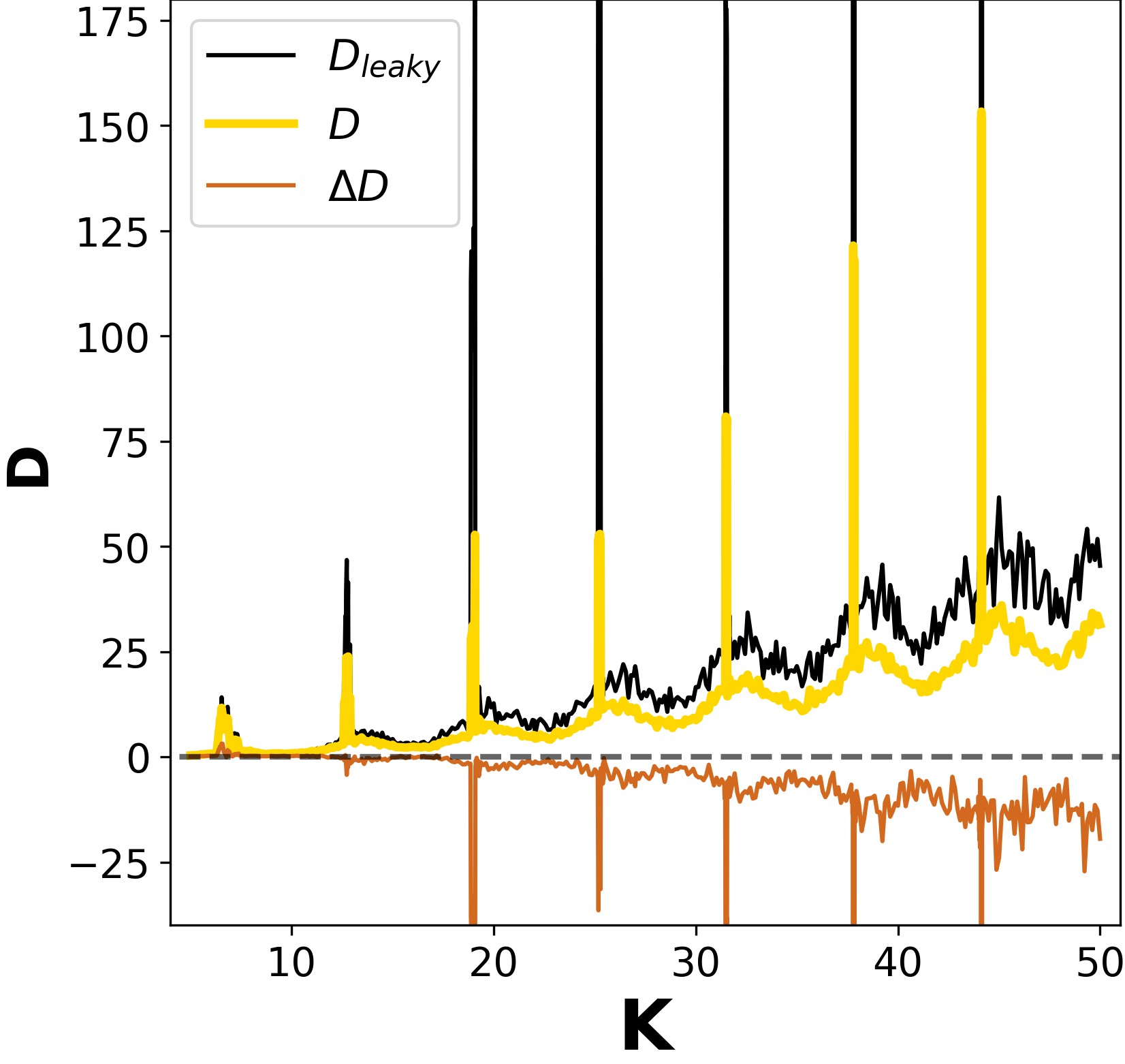}
    \caption{Diffusion coefficients in the open system (yellow), in the open-leaky system (black) \and the difference ($\Delta D$, brown) between them. Two holes are placed at $p=[10,10.5]$ and $[-10.5,-10].$}
	\label{deltaD}
\end{figure}

Let us begin to explore the difference between the open and the open-leaky system's diffusion coefficient $D$. To do this two holes are symmetrically defined in the interval of $p=[10,10.5]$ and $p=[-10.5,-10]$. Fig.~\ref{deltaD} depicts $D(K)$ curves for 500 different values in the interval $K=[5,50].$ As one can see, the diffusion coefficient can have very large values. The spiky structure, discussed thoroughly in the literature, is associated with accelerator modes in the phase space. According to the value of $K$ in Eq.~\eqref{acc_eq} the particles are pulled far away from their origin along the $p$ direction. It is worth mentioning that the shape of the $D(K)$ curve  does not depend on the length of the iteration. Taking $500$ time steps is sufficient to calculate $D(K)$. The only change that appears with longer simulation is that the spikes are getting higher because the accelerator modes take particles further and further. We can notice the similarity of the $D(K)$ curves for the open system ($D$), the open-leaky system ($D_{leaky}$ ), and also the difference between them ($\Delta D={D-D_{leaky}}$).
In case of a leaky system, $D_{leaky}$ consists of those particles only that survived even the last iteration step of the simulation. Consequently, negative values of $\Delta D$ mean that after 500 iterations the particles still in the system are further in the phase space on average. In other words, trajectories fallen out thorough the leak do not contribute to the diffusion coefficient. Resulting in a positive shift to the $D(K)$ curve. That is, we can point out that the main difference arises from the fact that the size of the ensemble is not constant during the simulation.

The question then naturally arises what does the $\Delta D$ depend on.

(i) \textit{initial conditions:} The uniformly distributed initial conditions as well as randomly chosen $(\Theta_0,p_0)$ pairs in the domain $\Theta=[0;1],\; p=[0;1]$ lead to the same result. 

(ii) \textit{leak size:} It turns out that the size largely contribute to the difference quantitatively, however, in a trivial manner.  We checked $\Delta D$ with several different leak size in $p$-direction, from $0.1$ to $1$ and found that the larger the leak, the bigger the difference $\Delta D.$ The reason behind this observation is simple: if we have a more elongated domain of escape, more particles run into it, thus, they enhance the contrast between the open and open-leaky diffusion.

\begin{figure}[htb]
    \centering
    \includegraphics[width=\columnwidth] {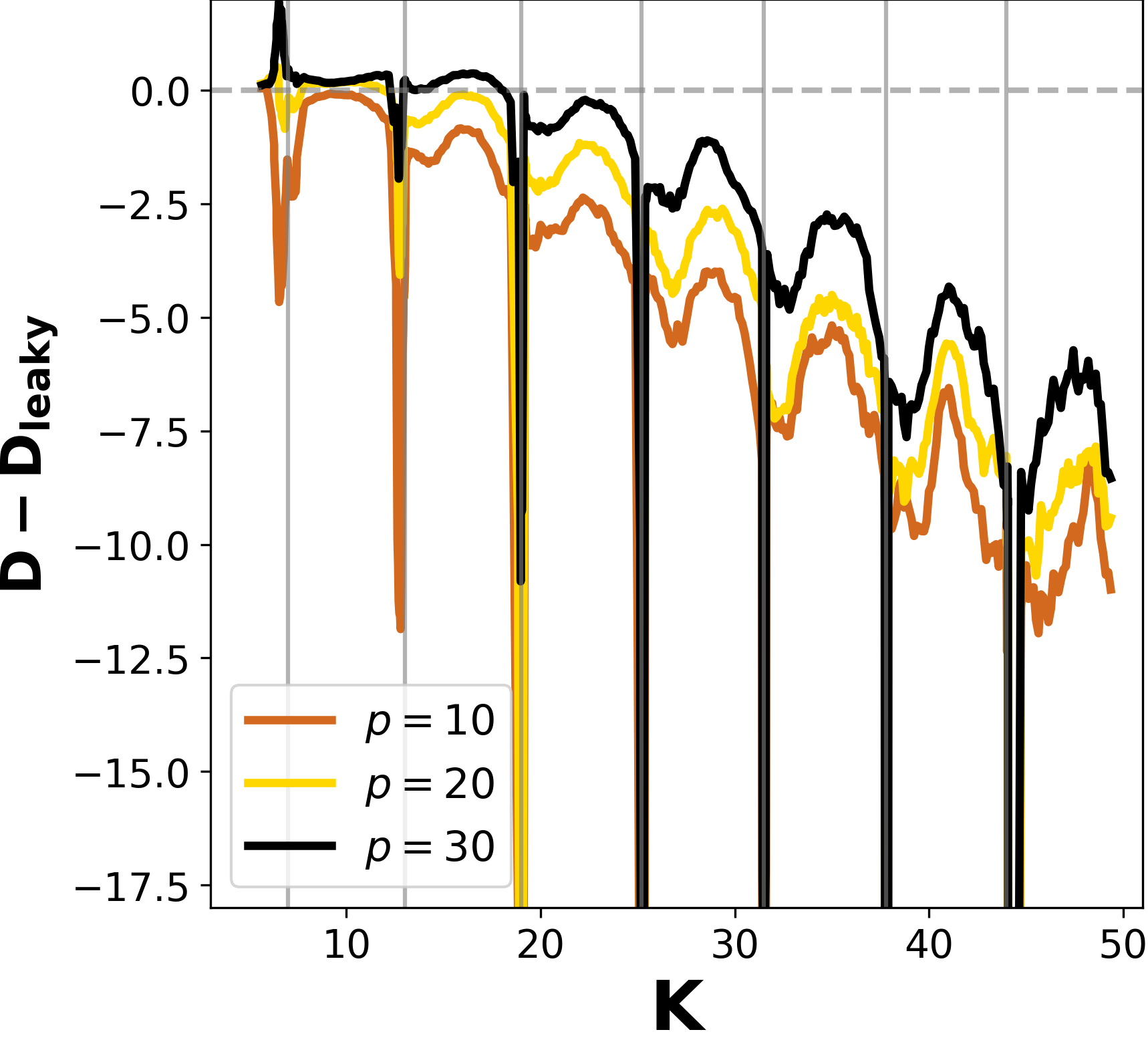}
    \caption{ For fix positioned holes at $p=[10,10.5]$ and $[-10.5,-10]$, the $D-D_{leaky}$ follows a decreasing trend (with peaks where the diffusion is anomalous) increasing the nonlinearity parameter. Also further placed leak does not effect on $\Delta D$ as much as if we put it closer.}
	\label{3pos_deltaD}
\end{figure}

(iii) \textit{leak position:} Repeating the previous simulation with leaks same in size ($\Delta p=0.5$) but at different positions along the $p$ axis: $p=20,-20$ or $30,-30$ we obtain $\Delta D(K)$ in Fig.~\ref{3pos_deltaD}. The brown curve coincides with $\Delta D$ in Figure~\ref{deltaD}. The upper two curves belong to leaks at $p=\pm20$ and $\pm30,$ respectively. The further we put the leaks to the initial conditions, the smaller difference appears in the diffusion coefficients. The explanation of this finding can be understood by examining the relative position and the initial ensemble position.

\begin{figure}[htb]
    \centering
    \includegraphics[width=\columnwidth] {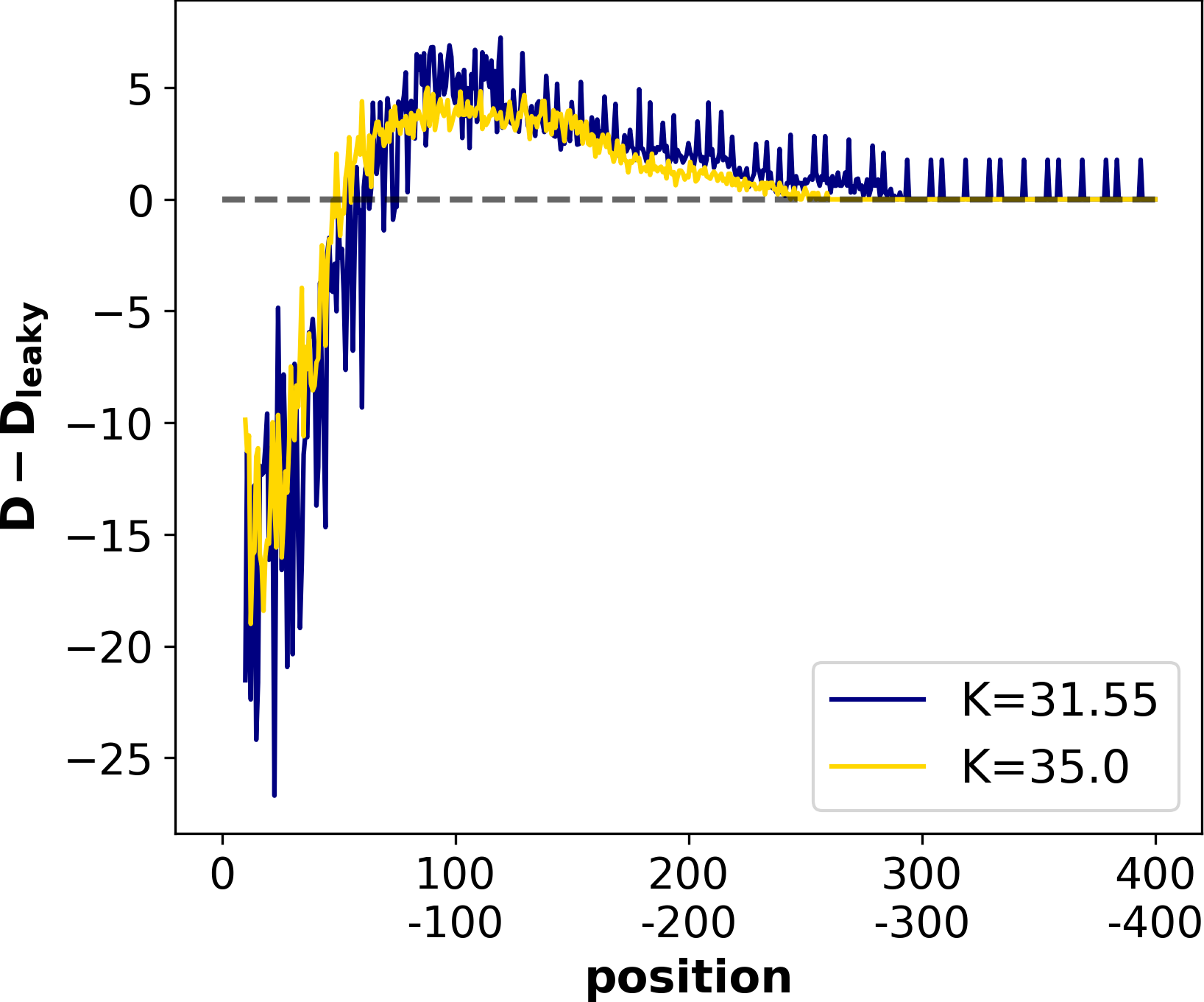}
    \caption{Position dependence of the $\Delta D$. At $K=31.55$, the accelerator modes in the phase space cause larger fluctuations than in the $K=35$ case.}
	\label{pos}
\end{figure}

We can also investigate the effect of leaks' position to the diffusion coefficient from a different aspect. Fig~\ref{pos} depicts $\Delta D$ for a fixed $K$ as a function of the leaks' position along the p-axis. The two curves stand for $K=31.55$ and $35$, the former one belongs to an accelerator mode in the phase space. 

The right tale of $\Delta D(K)$ corresponds to distant leaks. Here the difference of the two diffusion coefficients, $D-D_{\mathrm{leaky}},$ tends to zero because most of the particles experience normal diffusion and could not reach the leaks during the integration time. There are only a few of them stuck with AM islands. If these trajectories enter the leaks, the almost periodic spikes are appearing on top of the (blue) curve yielding a significant difference in $\Delta D.$

On the other hand, for closer leaks, left side in Figure~\ref{pos}, we have negative values according to Figure~\ref{3pos_deltaD} as described above.

There is also a special leak position for every $K$ being in harmony with the expectation value of squared distance $(p-p_0)^2$. That is, the intersection with horizontal axis, i.e. where $\Delta D=0,$ describes the situation when the leaks engulf the trajectories that provide the displacement from their origin approximately equal to $D$ in non-leaky system. Beyond this point, the leaks capture the faster trajectories and, therefore, the survivors produce smaller diffusion coefficient ($D_{\mathrm{leaky}}$) than $D$ resulting in a positive deviation.
The maximum of $\Delta D(K)$ can be thought as a barrier when escape takes place at the ''wavefront'' of normal diffusion removing the ''fastest'' particles from the dynamics leaving behind the largest gap between leaky and non-leaky scenarios.

In Fig~\ref{pos}, taking a deeper look at the trend of the curves from right to left we can recognize the similarity with the envelope of the curves in Fig~\ref{3pos_deltaD}. If we consider a pair of leaks at fix position and increase $K$ (keeping the integration time also fixed), we observe the spreading of the particles along $p$ axis. The relative size of the interval between the leaks and the spanned interval by the particles movement decreases with larger $K$. For fix $K$, the result in the phase space is the same if we place the leaks closer to each other.

\subsection{Survival probability, escape rate}
\begin{figure}[htb]
    \centering
    \includegraphics[width=\columnwidth] {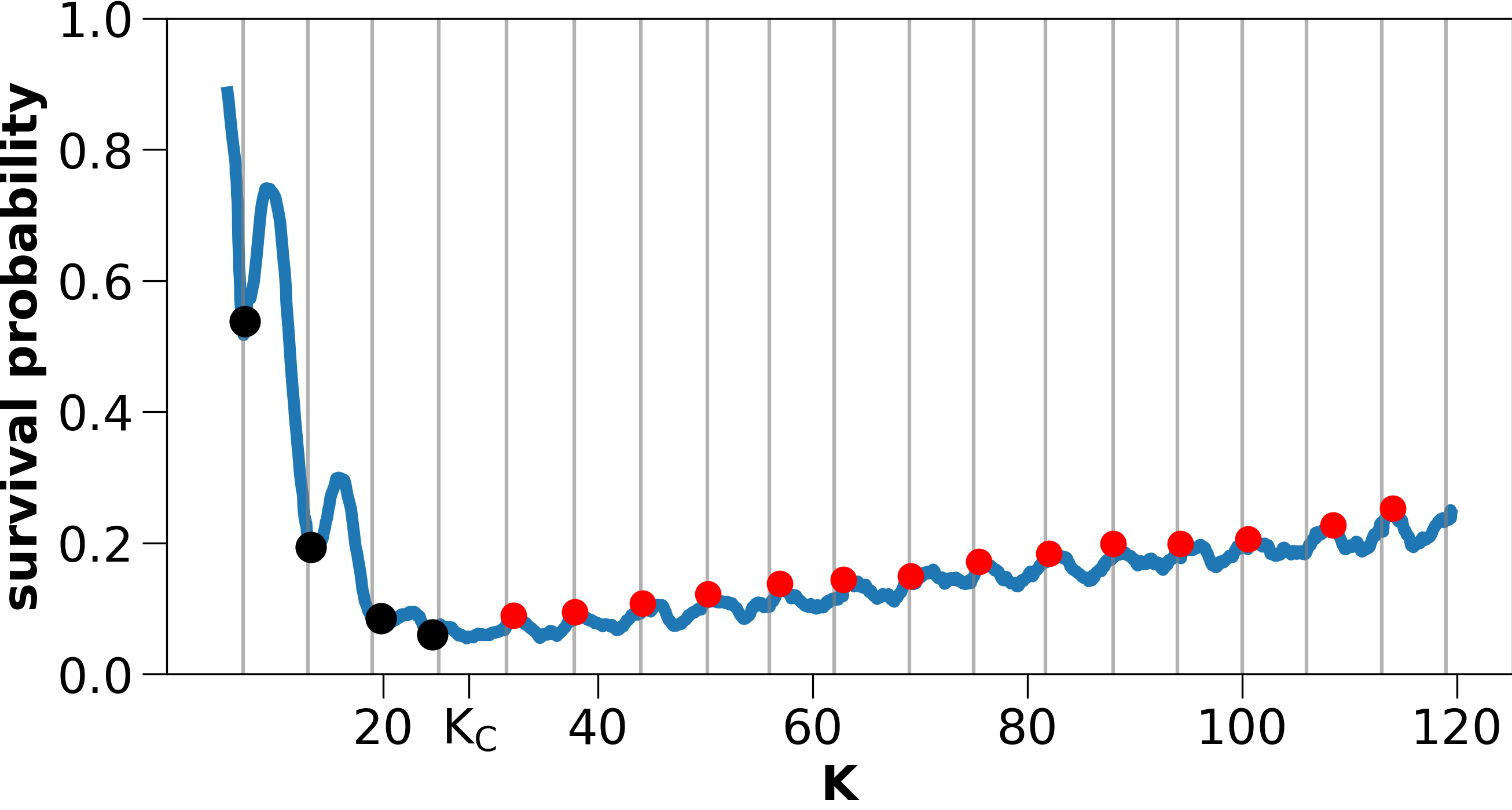}
    \caption{Survival probability for particles in a leaky system. Vertical lines mark the presence of accelerator mode for that $K$. The holes are at $p=30$ and $-30$ with height $0.5$. For small $K$ values, accelerator modes increase the chance of leaving and the survival probability curve has local minima (black circles) there. However, at large $K$, they keep particles in the system, therefore,  the possibility of staying in the system is higher (red circles).}
	\label{survival_prob}
\end{figure}

Up to now, we learned about diffusion coefficient in open-leaky and its relation to the open standard map. Basically, the survived particles govern the $\Delta D$, therefore, it is worth investigating the escape of particles through the predefined leaks. At this point we turn our focus of interest to the number of particles left the system.  More precisely, the ratio of the particles still in the system and the initial population of the ensemble as time goes by is measured numerically. To do this, the leaks are specified symmetrically at $p=30$ and $-30$ with height $0.5.$ Figure~\ref{survival_prob} portrays the evidence that for small $K$ ($<10$) values low amount of the particles manage to reach the leaks. Consequently, the survival probability is high $P\gtrapprox 0.5$. As we increase the non-linearity parameter they can get to the leaks more likely since the stronger diffusion drives them further. This is how the decreasing part of the curve is explained. The minimum of the survival probability at $K_c$ stands for the case when most of the particles cover the distance to the holes and eventually escape. The ascending trend, then, is annotated by jumping over the leak due to the stronger diffusion yielding extended loops in $p$, in one iteration step. The role of the accelerator modes appears along the vertical gray lines as local minima (black circles) and maxima (red circles) in the decreasing and increasing parts, respectively. As we know, accelerator modes help particles travel further in the same amount of time. Thus, they assist the particles to reach the leaks in the low $K<K_c$ regime, while help them jump over the leak and end up far distance for large values, $K>K_c$. This is why we observe local minima (maxima) at accelerator modes before (after) a critical $K_c$ value of nonlinearity parameter.

In order to talk about escape rate, we have to fix the parameters of the simulations. It is clear that we need more particles to have reliable statistics, therefore, we choose to follow $10^6$ trajectories. We also raise the length of the iteration to $10^7$ in order to be able to analyze the tail of the distribution. The leaks are now settled at $p=[-10.5,-10]$ and $[10,10.5]$.

\begin{figure}[htb]
    \centering
    \includegraphics[width=0.95\columnwidth] {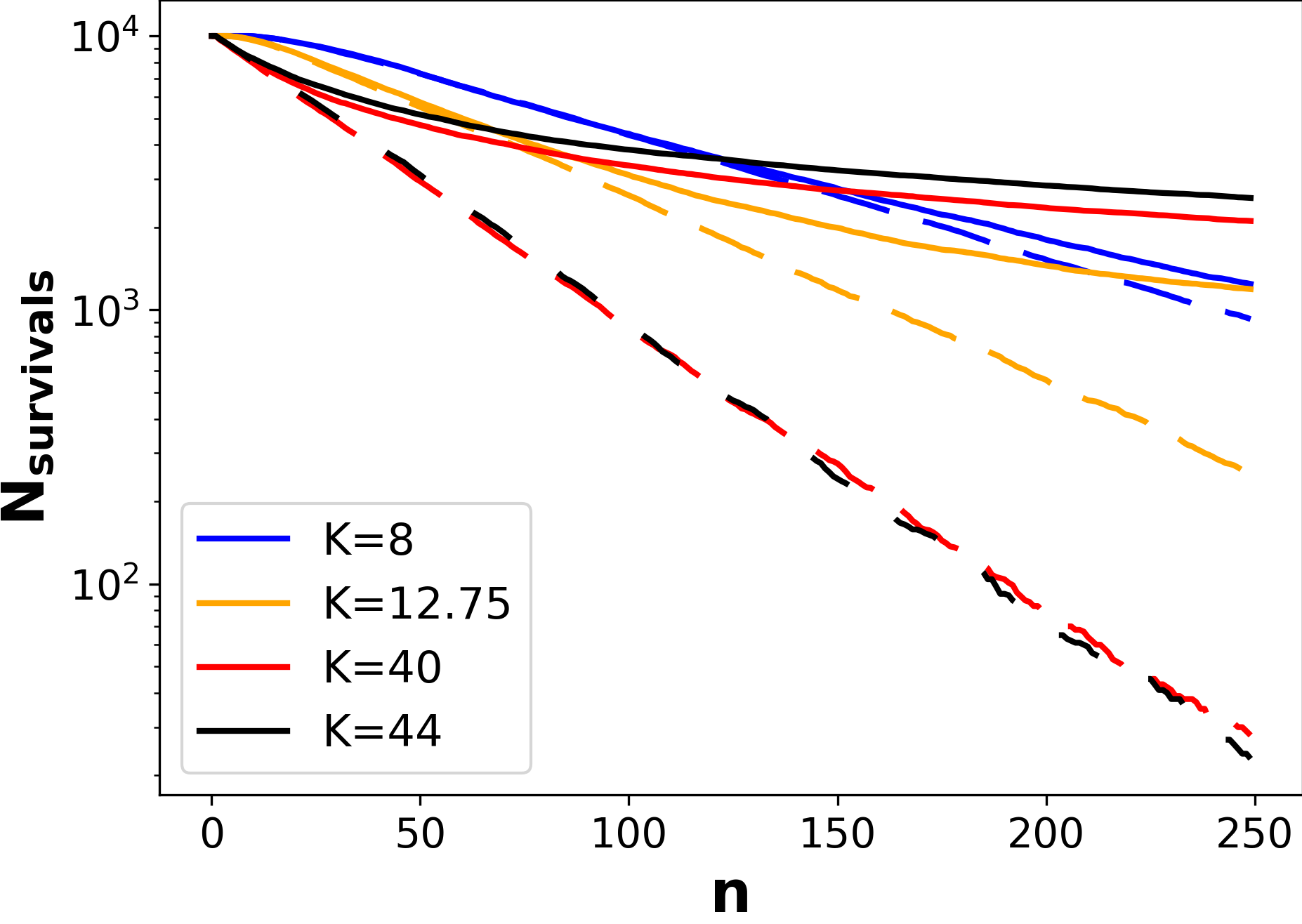}
    \caption{Number of survived particles for different $K$ values for short times does not follow exponential distribution for open-leaky standard map (solid lines) and true exponential for close-leaky standard map (dashed lines) limited to $p=[-40,40]$ via periodic boundary conditions. Different slope of the dashed lines - the escape rate - characterizes the quality of mixing. The larger the $K$, the shorter the recurrence time to the leaks.}
	\label{short_escape}
\end{figure}

Let us begin with a classical simulation. Initial conditions are placed uniformly in a unit square ($p_0,\Theta_0$)=[-0.5;0.5]x[0;1]. The leaks cover the area mentioned above. The number of non-escaped trajectories vs. time is stored. We stress that the smallest value of $K$ is far beyond the non-linearity parameter that generates fully ergodic dynamics. Thus, without a doubt, we assume a clear exponential trend based on the literature. The results are summarized in Figure~\ref{short_escape}. Solid lines correspond to leaky-open standard map and clearly deviates from a straight line in the log-lin plot. This observation completely contradicts our expectation, and not only for accelerator modes ($K=12.74$ and $44$) but for normal diffusion ($K=8$ and $40$) as well.

In our study the motion can be considered on a cylinder, $p\in (-\infty, \infty), \Theta\in [0,1].$ So, periodic boundary condition in $\Theta$ allows that a trajectory re-enters the same domain in the phase space frequently. This view in $p$-direction is more contrasting. Due to the unbounded motion in momentum the trajectories might wander much longer times in phase space before they come back to the same position. To check what happens if we divert back the particles violently to the same realm in phase space, we repeat the previous process with periodic boundary condition, $p$ $\mod(40)$. The resulting dashed curves in Figure~\ref{short_escape} follow the desired exponential decay. As the only difference between the two simulations is the open or closed phase space, we conclude that the non-exponential decay for open-leaky system can be a consequence of the unbounded phase space in $p$ coordinate. Periodic boundary conditions do not let particles leave the surrounding area of the leaks, hence, the well-mixing process, which is the  basic criterion of the Poisson survival distribution for short times \cite{Sinai1966,Benkadda1994,Zaslavsky1997a}, is violated.

Interestingly, one can observe the same behavior if $p=\pm 40$ behaves as a strict edge of the system, i.e. particles leave the system, ''fall out'', when their momentum exceeds $|p|\geq 40.$ This phenomenon has also been reported in  \cite{Livorati}.
\begin{figure}[htb]
    \centering
    \includegraphics[width=0.95\columnwidth] {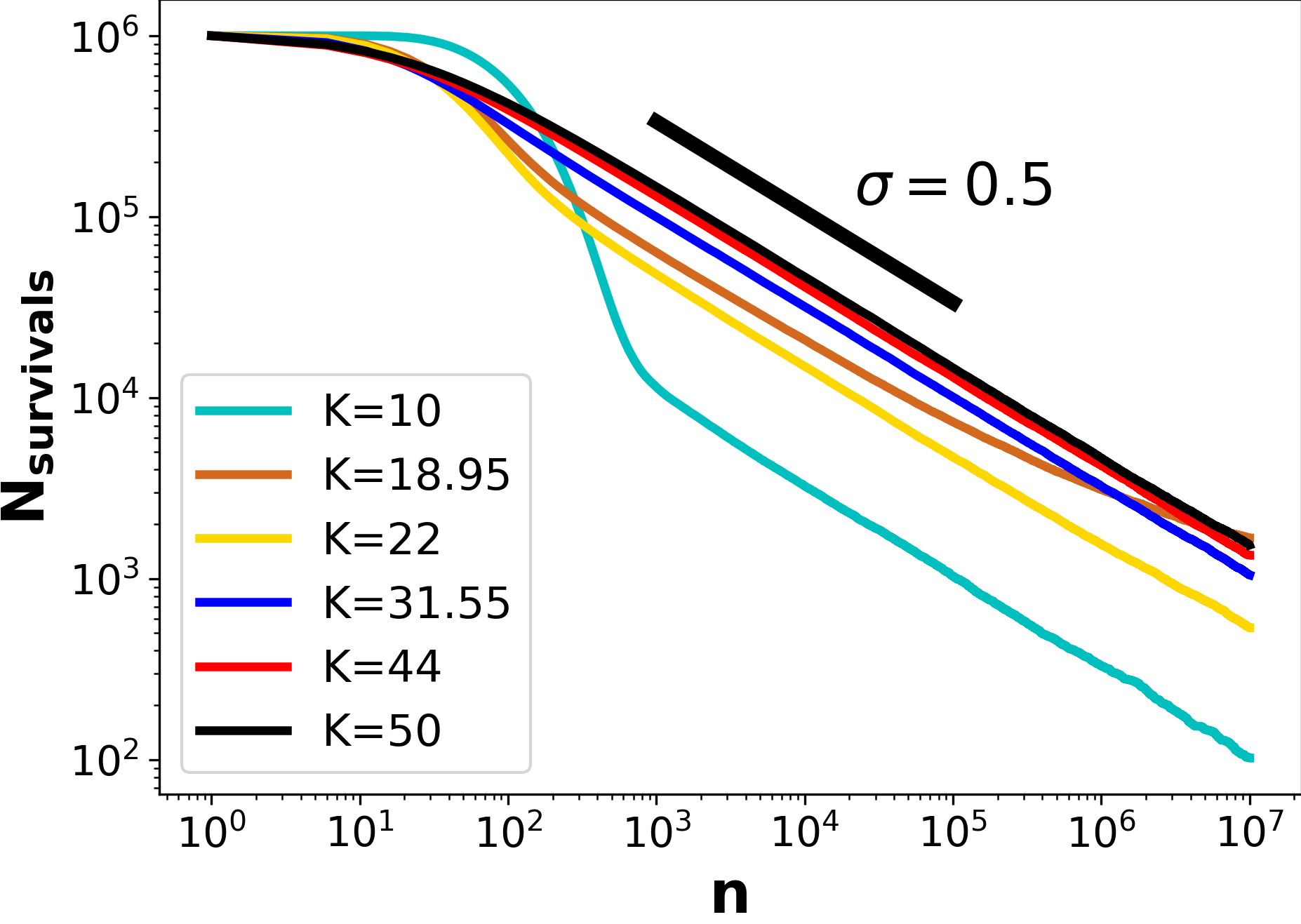}
    \caption{For different $K$ values the number of survived particles until $n=10^7$. The decays follow power-law for each case. The slopes of the curves are nearly the same $\sigma\approx0.5$. }
	\label{long_escape}
\end{figure}

The long-term dynamics in open-leaky problem fits to our presumption. Figure~\ref{long_escape} represents the power-law tail of survival distribution up to $10^7$ iteration steps for different $K$ values. The tail ($n>1000$) of various $N(t)$ has nearly the same slope, $\sigma=0.5$ Only the cyan curve corresponding to the AM $K=18.95$ shows a deviation beyond $n\approx 3\times 10^5$. 
Moreover, the clear exponential part is also visible, especially in the case of $K=10.$ From this and Fig.~\ref{short_escape}, we can point out that the initial stage of the survival distribution ($n\lesssim 200-300$) indicates exponential decay only for moderate control parameters. This fact reinforces the view that larger diffusion works against the efficient mixing in phase space.

\begin{figure}[htb]
    \centering
    \includegraphics[width=0.95\columnwidth] {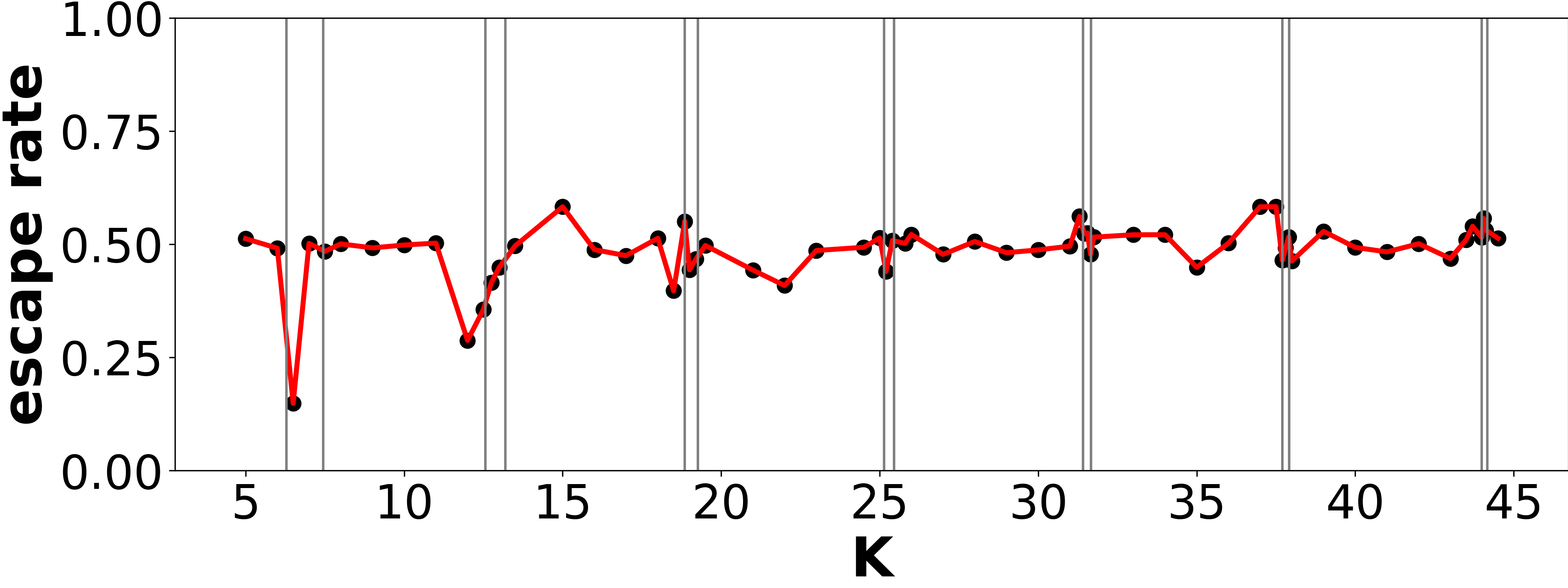}
    \caption{Escape rate for different $K$ parameters. The fit was executed after $5000$ iterations where the $N(n)$ curve was straight line on log-log scale.}
	\label{exponent}
\end{figure}

Although we found that after ca. 5000 iterations, the curves in Fig.~\ref{long_escape} become linear on log-log scale, it should be noted that the escape rates are not exactly equal. Fig.~\ref{exponent} collects  the power-law exponents, $\sigma$, for 60 different $K$ values. The fit has been made on straight segment of the distributions. It turns out from the fluctuations of the $\sigma(K)$ plot that the $N(t)\sim1/\sqrt{t}$ rule does not hold for the accelerator modes, for instance, $6.28 < K < 7.45$ or $12.57 < K < 13.19$. There are, however, shorter sections of $N(t)$ curves at AMs that follow the $\sigma=0.5$ exponent but their tail always deviates from the straight line causing different slope. It cannot be rule out that for longer iteration, say, $10^9-10^{10}$ steps, these curves pursue again the original slope or tend to different value of $\sigma.$ This investigation is beyond the scope of present work.

\begin{figure}[htb]
    \centering
    \includegraphics[width=\columnwidth] {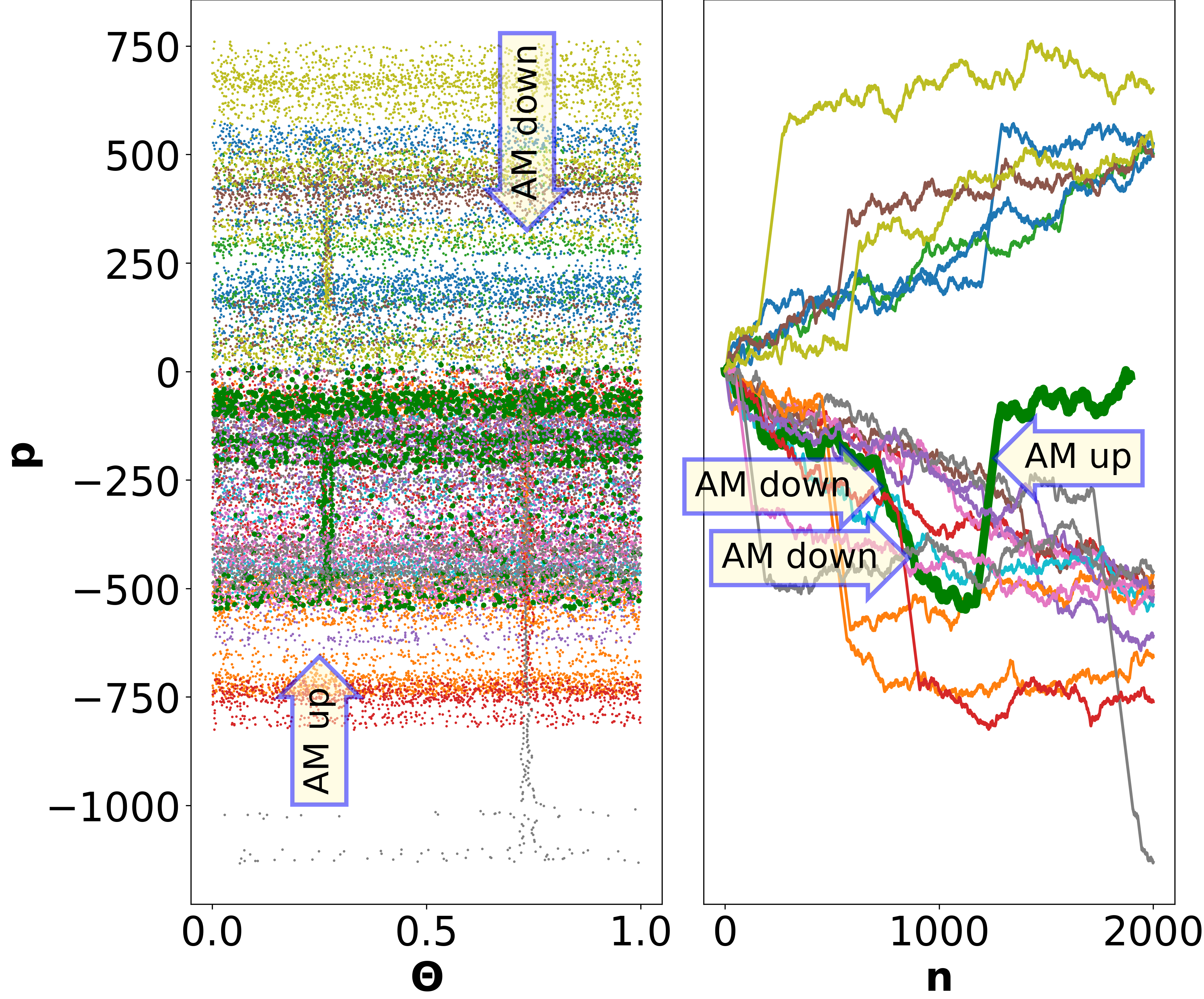}
    \caption{Left: Phase space portrait ($K=25.3$) with accelerator (decelerator) modes. Right: Momenta of 19 trajectories experiencing AMs. The green curve shows a particle travelling downward then trapped by the left accelerator channel -- that acts as a decelerator mode in lower semi-plane --  and finally enters backward the bottom leak around $n\approx 1900$.}
	\label{fig:backward_escape}
\end{figure}

\begin{figure}[htb]
    \centering
    \includegraphics[width=\columnwidth] {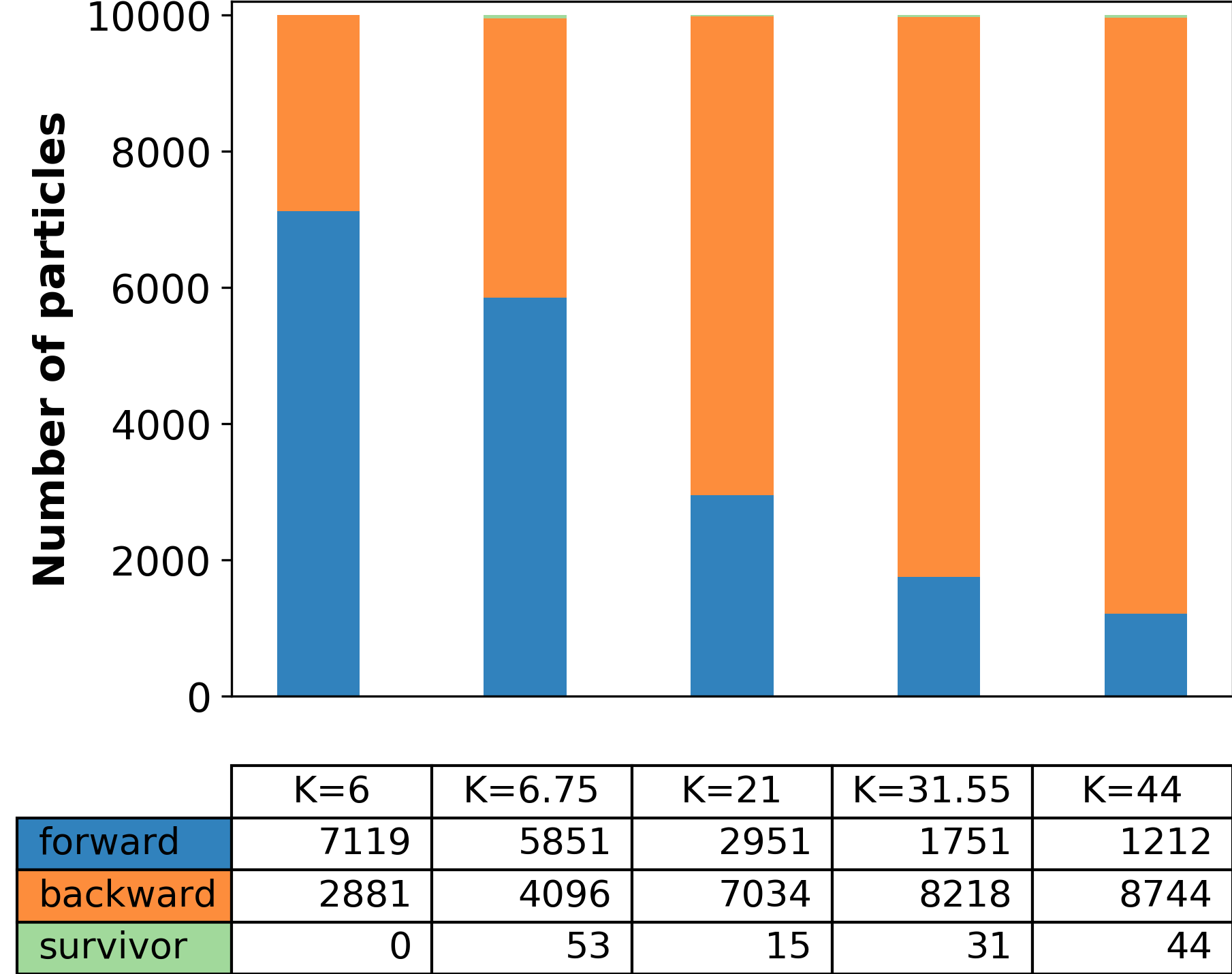}
    \caption{Particles classified to three categories according to the "direction of leaving" for five different $K$ values after $10^6$ iterations. Increasing $K$ leads to more particles escape during backward motion. The group of survivors is not visible on top of the bars.}
	\label{bar_chart}
\end{figure}

The finite size of the leaks make it possible for the particles to jump over them and get to larger $p$ coordinates. Indeed, some of them are not so ''lucky'', mostly for small $K$, and their furthest position to the origin is situated exactly in the leak. The others, however, can enter the leaks while they change the direction of movement and return to smaller values of $p$, see Figure~\ref{fig:backward_escape}.

As they have their momentum $p$ larger (smaller) than the leak's upper (lower) border, let us call them \textit{backward} particles and the former category of escaping particles \textit{forward.} Of course, there are particles that do not leave the system, they are the \textit{survivors}. In Fig.~\ref{bar_chart} we illustrated the number of particles in the three groups for five different values of $K$ after $10^6$ iterations. One can see that for larger $K$ more particles leave the system backward. 

It is not surprising, since they have more chance to jump over the leaks. Comparing the first two columns, the significant difference in the ratio of the forward and backward particles is remarkable. A tiny increase in $K,$ from 6 to 6.75 refers to spike structure of Figure~\ref{deltaD} and it is the consequence of an appearing accelerator mode around $K=6.5$.

\begin{figure}[htb]
    \centering
    \includegraphics[width=\columnwidth] {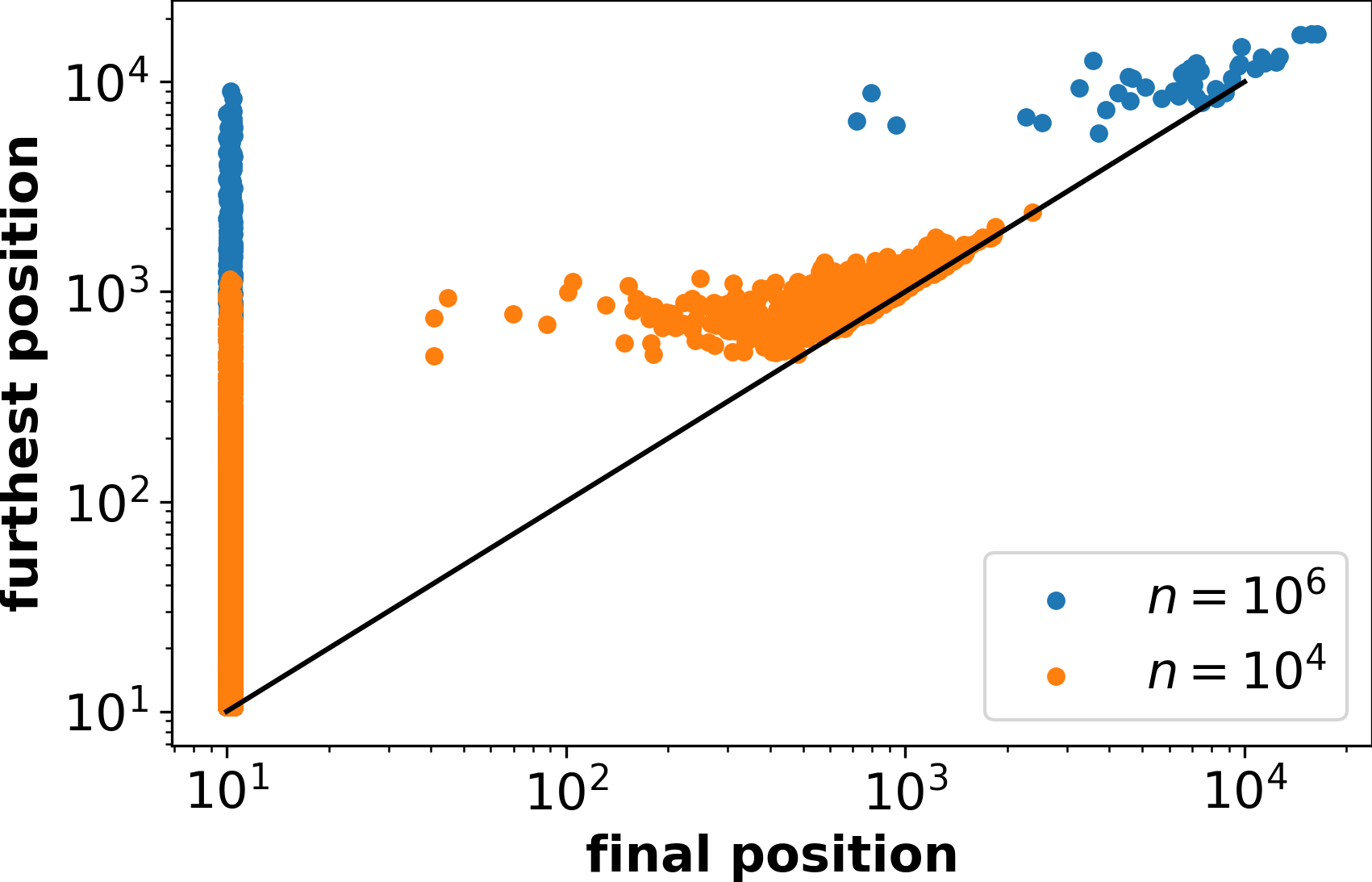}
    \caption{For $K=50$ absolute value of the $p$ coordinate  at the furthest position and at the last iteration for $n=10^4$ (orange) and $n=10^6$ (blue). The final position is $0$ if a particles not left the system. Considering longer times make the scale change and more particles end up leaving via the leaks.}
	\label{pos_max}
\end{figure}

We have already seen that not all of the particles leave the system while they approximate one of the leaks. In  Figure~\ref{pos_max}, we depicted the ensemble at two particular time instants: the absolute value of the largest $p$ coordinate the particles reached vs. the absolute value of $p$ at the last iteration step before leaving. Whenever a particle escapes, its final momentum is set to the last position in $p$ (it belongs obviously to either leaks). The vertical blue/orange lines at $\approx 10$ indicates the position of the escaped particles right before falling into the leak. One can see that the remaining points are scattered above the $y=x$ line that means they travelled far away in the phase space. A considerable amount of iteration increases the number of escaped particles and might increase, with the help of AMs and their stickiness effect, the distance from the initial place for the survivors. If we compare the two calculations, $n=10^4$ (orange points) and $n=10^6,$ (blue) we can conclude that every further iteration, particles might escape or get further in the phase space. This can be considered as the main fingerprint of power-law decay in particles long-time distribution.

\section{Summary and conclusion}
\label{conclusion}
In this paper, we obtained the diffusion coefficient and escape rate in the open standard map ($p\in(-\infty,+\infty)$) with two finite size holes placed symmetrically to the origin along the $p$ coordinate. Our main goal was to explore the difference between the diffusion coefficients in the presence of artificial leaks in the system.

Based on numerical investigation of large ensembles of particles in the phase space, we found the followings.

First of all, it becomes clear that punching the phase space by artificial holes results difference in diffusion coefficients in leaky and non-leaky version of open standard map.  This can be understood from the fact that disappearing particles have direct influence to the diffusion coefficient $D$ via square of the displacement in $p-p_0.$ We have also shown that for a given non-linearity parameter the relative position and size of the leaks influences significantly $\Delta D=D-D_{\mathrm{leaky}}.$  That is, leaks closer to the initial conditions or with larger size have a remarkable impact on the $\Delta D.$ These statements are strongly correlated to the parameter $K.$ Since  either in normal or in acceleration modes both $D$ and $D_{\mathrm{leaky}}$ depends on $K.$

Since the escaping particles play an important role in diffusion coefficients, we also investigated the escape dynamics for short and long timescales.
We found it surprising that the distribution of the survived particles for short times is not exponential as expected for strong chaos. We believe that the reason of this is none other than that in cylindrical phase space without periodic boundary conditions in momentum the particles can take large excursions in the phase space before they come back again to the vicinity of the leaks and escape. This behavior implies that the perfect mixing is not fulfilled although it is required to observe the Poisson distribution in escape time statistics.

A more classical result corresponding to the long time escape statistics has been achieved. That is, the well-known power-law decay of particles with nearly identical exponent ($\sigma=0.5$) is manifested in open-leaky SM. Nevertheless, the  escape rate also depends on whether accelerator modes are operating in phase space.

Additionally, we presented the fact that direct escape (forward particles) is common only for low values of $K$ and realized that quite a large amount of particles, especially for large K values, travel far in the phase space before they fall into one of the leaks (backward particles). As the backward particles can reach very larger distances to their origin in $p$ co-ordinates, we might have to wait for them to come back and escape. That is why a power-law distribution describes the escape rate.

The universality of the standard map allows us that the results established in this study can be applied in other fields of science where the open-leaky properties of the phase space are relevant, such as hydrodynamics, environmental flows, chemical reactions, or formation and evolution of planetary systems.

\begin{acknowledgments}
This work was supported by the NKFIH Hungarian Grants K119993, K125171, FK134203. The support of Bolyai Research Fellowship and \'UNKP-19-4 New National Excellence Program of Ministry for Innovation and Technology is also acknowledged.
\end{acknowledgments}

\bibliography{references}
\end{document}